\documentstyle[12pt]{article}
\newlength{\extraspace}
\newlength{\extraspaces}
\newcommand{\be}{\begin{equation}
\addtolength{\abovedisplayskip}{\extraspaces}
\addtolength{\belowdisplayskip}{\extraspaces}
\addtolength{\abovedisplayshortskip}{\extraspace}
\addtolength{\belowdisplayshortskip}{\extraspace}}
\newcommand{\ee}{\end{equation}}

\newcommand{\ba}{\begin{eqnarray}
\addtolength{\abovedisplayskip}{\extraspaces}
\addtolength{\belowdisplayskip}{\extraspaces}
\addtolength{\abovedisplayshortskip}{\extraspace}
\addtolength{\belowdisplayshortskip}{\extraspace}}
\newcommand{\ea}{\end{eqnarray}}

\newcommand{\bas}{\begin{eqnarray*}
\addtolength{\abovedisplayskip}{\extraspaces}
\addtolength{\belowdisplayskip}{\extraspaces}
\addtolength{\abovedisplayshortskip}{\extraspace}
\addtolength{\belowdisplayshortskip}{\extraspace}}
\newcommand{\eas}{\end{eqnarray*}}

\newenvironment{theorem}[1]
{\vspace{3mm}\noindent {\bf #1:}}%
{\vspace{2mm}}
\newcommand{\bt}[1]{\begin{theorem}{#1}}
\newcommand{\et}{\end{theorem}}

\newcommand{\newsection}[1]{
\vspace{15mm}
\pagebreak[3]

\addtocounter{section}{1}
\setcounter{subsection}{0}
\setcounter{footnote}{0}
\begin{flushleft}
{\large\bf \thesection. #1}
\end{flushleft}
\nopagebreak
\medskip
\nopagebreak}

\newcounter{prop}

\newcommand{\N}{\mbox{I\hspace{-.4ex}N}}
\newcommand{\C}{\,\mbox{{\sf I}\hspace{-1.2ex}{\bf C}}}
\newcommand{\Z}{\mbox{Z\hspace{-.8ex}Z}}

\newcommand{\bra}{\langle\,}
\newcommand{\ket}{\,\rangle}
\newcommand{\sspace}{\makebox[1cm]{ }}
\newcommand{\bspace}{\makebox[2cm]{ }}
\newcommand{\nspace}{\!\!\!\!\!\!\!\!\!\!}
\newcommand{\is}{\! & \! = \! & \!}
\newcommand{\nonum}{\nonumber \\[1.5mm]}
\newcommand{\ra}{\rightarrow}

\newcommand{\rra}{\ \longrightarrow \ }

\renewcommand{\d}{\partial}
\newcommand{\dd}[1]{{\partial \over \partial {#1}}}

\newcommand{\Tr}{{\rm Tr}}

\newcommand{\Ker}{{\rm Ker\,}}

\newcommand{\lb}{\lambda}
\newcommand{\A}[1]{{ A_{N,#1} }}
\newcommand{\B}[1]{{ B_{N,#1} }}
\newcommand{\Ab}[1]{{ \overline{A}_{N,#1} }}
\newcommand{\Vb}[1]{ \overline{V}_{#1} }
\newcommand{\Ao}[1]{{ A_{N,#1}^{(0)} }}
\newcommand{\Bo}[1]{{ B_{N,#1}^{(0)} }}

\newcommand{\Vbo}[1]{ \overline{V}_{#1}^{(0)} }
\newcommand{\LA}[1]{{ L(A_{N,#1}) }}
\newcommand{\LB}[1]{{ L(B_{N,#1}) }}
\newcommand{\LV}[1]{{ L(V_{#1}) }}
\newcommand{\LVb}[1]{{ L(\overline{V}_{#1}) }}
\newcommand{\LAb}[1]{{ L(\overline{A}_{N,#1})}}
\newcommand{\LAo}[1]{{ L(A_{N,#1}^{(0)}) }}
\newcommand{\LBo}[1]{{ L(B_{N,#1}^{(0)}) }}

\newcommand{\LVbo}[1]{{ L(\overline{V}_{#1}^{(0)}) }}
\newcommand{\cH}{{\cal H}}
\newcommand{\hH}{\widehat{{\cal H}}}
\newcommand{\Ch}{\widehat{C}}
\newcommand{\cF}{{\cal F}}
\newcommand{\gh}{{\widehat{g}}}
\newcommand{\cA}{{\cal A }_N}
\newcommand{\cAb}{\overline{{\cal A}}_N}
\newcommand{\cB}{{\cal B }_N}
\begin{document}
%
%%%%%%%%%%%%%%%%%%%%%%%%%%%%%%%%%%%%%%%%%%%%%%%%%%%%%%%%%%%%
\addtolength{\textwidth}{-1cm}
\begin{titlepage}
\begin{flushright}
MPI-Ph/93-91\\
November 1993
\end{flushright}
\vspace{1cm}

\begin{center}
{\LARGE   Infinite Abelian Subalgebras\\
            in Quantum W-Algebras:\\[3mm]
          $\;\;\;$ An Elementary Proof}
{\makebox[1cm]{ }
          \\[2cm]
{\large Max R. Niedermaier}\\ [3mm]
{\small\sl Max-Planck-%
Institut f\"{u}r Physik} \\
{-\small\sl Werner Heisenberg Institut - } \\
{\small\sl 80805 Munich (Fed. Rep. Germany)}}
\vspace{3.5cm}

{\bf Abstract}
\end{center}
\begin{quote}
An elementary proof is given for the existence of
infinite dimensional abelian subalgebras in quantum
W-algebras. In suitable realizations these subalgebras
define the conserved charges of various quantum
integrable systems. We consider all principle W-algebras
associated with the simple Lie algebras. The proof is
based on the more general result that for a class of
vertex operators the quantum operators are related to
their classical counterparts by an equivalence transformation.
\end{quote}

\vfill
\end{titlepage}

%
%%%%%%%%%%%%%%%%%%%%%%%%%%%%%%%%%%%%%%%%%%%%%%%%%%%%%%%%%%%
\newsection{Introduction}
For each simple Lie algebra $g$ there is an associated
principle $W$-algebra $W(g)$, defined in its free field
realization through the intersection of the kernels of
screening operators. On the classical level the structure
of these $W$-algebras can be investigated by various
techniques -- most of which, however, fail to have
counterparts in the quantum theory. In particular, the
mere construction of a quantum $W$-algebra requires novel
techniques\cite{Lu,MN2,FF2}. Conceptually, their existence
can be understood by treating the $W$-algebra
as a vertex operator algebra whose state space
appears as the only non-vanishing cohomology class of
a Fock space resolution. Either directly \cite{MN2}
(for $sl(n)$) or with reference to the classical limit
\cite{FF2} this proves the existence of a quantum
$W$-algebra $W(g)$ associated with each simple finite
dimensional Lie algebra $g$.

An important feature of these quantum $W(g)$-algebras
is the conjectured existence of infinite dimensional
abelian subalgebras $I(\gh)$ labeled by an affine
Lie algebra $\widehat{g}$ associated with $g$. For
$\widehat{g}=\widehat{sl}(2)$ this conjecture originally
arose in A.B. Zamolodchikov's program of perturbing a
conformal field theory to obtain integrable massive
quantum field theories; the generators of the abelian
subalgebra playing the role of the conserved charges,
which enforce the factorized scattering theory. In
suitable realizations the generators of $I(\gh)$ can be
constructed in terms of rank$(g)$ scalar fields, which
may be free or interacting. In the latter case the
generators play the role of the conserved charges
in various quantum integrable systems (affine Toda,
mKDV, KdV). The conserved charges can be characterized
as integrals over differential polynomials in the scalar
fields which lie in the kernel of some operator $ad\,Q_0$
-- the conjecture being that this kernel is infinite
dimensional.

The proof of this conjecture turned out to be surprisingly
difficult. Again a simple transcription of classical
techniques fails. In \cite{MN1} a constructive
proof was given for the low members of the $A_r^{(1)}$
series. Conceptually one expects that a
suitable extension of the cohomological construction
of quantum $W$-algebras would also allow one to
characterize their infinite abelian subalgebras.
The correct complex to achieve this was identified by
B. Feigin and E. Frenkel \cite{FF1} as a spectral
sequence of Fock spaces whose cohomologies are
isomorphic to the exterior algebra
$\bigwedge^*(\widehat{a}^*)$ of the dual space to the
principle commutative subalgebra $\widehat{a}$ of
$\widehat{n}_+$. The proof is then based on the
stability of the Euler characteristic of the
complex under deformations of a parameter. Because
of the alternating sum, however, the argument goes through
only for affine Lie algebras whose exponents and
Coxeter number are odd and even, respectively\cite{FF3}.
This excludes $A_n^{(1)},\;n>1,\;
D_{2n}^{(1)},\;n>1,\;E_6^{(1)}$ and $E_7^{(1)}$.
Below we will give a proof by elementary techniques
which is valid for all affine Lie algebras.
In particular this covers the missing cases. The proof
rests on the more general result that the operator $Q_0$
is related to its classical counterpart $Q_0^{(0)}$
by an equivalence transformation.
\pagebreak
%%%%%%%%%%%%%%%%%%%%%%%%%%%%%%%%%%%%%%%%%%%%%%%%%%%%%%

\newsection{$W(g)$-modules and their classical limit}
{\em 2.1.} Let $g$ be a simple finite dimensional Lie
algebra of rank $r$, and consider $r$ copies of an
infinite dimensional Heisenberg algebra
\bas
[a_m^a,\,a_n^b]\is \beta^2 m\delta_{m+n,0}\delta^{ab}\;,
\sspace 1\leq a,b\leq r,\;\;m,n\in \Z\;,
\eas
where $\beta$ is a formal parameter with inverse $\beta^{-1}$
generating the field $\C[\beta]$. It is convenient to use
the realization
$a_{-m}^a=m x_m^a,\;a_m^a =\beta^2 \dd{x_m^a},\;m>0$.
Introduce Fock spaces labeled by elements $\lb$ in the
dual $h^*$ of the Cartan subalgebra of $g$ by
$\pi_0^{(0)}=\;\C[x_1^a,x_2^a,\ldots],\;\pi_{\lb}^{(0)}=
\pi_0^{(0)}\otimes\C v_{\lb}$ and define a shift
operator $T_{\lb}:\pi^{(0)}_{\lb'} \ra
\pi^{(0)}_{\lb+\lb'}$, $P\otimes v_{\lb}\ra P\otimes
v_{\lb +\lb'}$. We shall need also Fock spaces $\pi_{\lb}$
considered as modules over $\C[\beta^2]$. As linear spaces
set $\pi_{\lb}= \pi_{\lb}^{(0)} \otimes\, \C[\beta^2]$. Let
$C:\pi_{\lb}\ra \pi_{\lb}^{(0)}$ denote the projection
onto the $\beta$-independent (`classical') part. The
Fock spaces $\pi_{\lb}$ carry two natural graduations:
The principal graduation $\deg x_n^a =n,\;\deg v_{\lb}
=0= \deg \beta$ and the power graduation power$\,x_n^a=1$,
power$\,v_{\lb}=0=$ power$\,\beta$. Denote by
$(\pi_{\lb})_{N,m}$ the subspace of degree $N$ and power
$m$.

The central objects in the following are operators
of the form
\be
Q_{\lb}:\pi_0 \rra \pi_{\lb},\sspace \lb \in h^*\;,
\ee
defined by $Q_{\lb} = T_{\lb}\,(V_{\lb})_1$, where
\bas
&&\nspace  \beta^2 (V_{\lb})_k = \sum_{m\geq 0}
S_{m-k}\left[\{\lb\cdot a_{-n}/n\} \right]
S_{m}\left[\{-\lb\cdot a_n /n\} \right] \;.
\eas
$S_m[\{x_n\}],\;m\in \Z$ ($S_m=0,\; m<0$) are the elementary
Schur polynomials in $x_n,\; n>0$. It is often convenient to
decompose these operators w.r.t. the powers of the
annihilation operators appearing. A term of power $N$ in
$\lb\cdot a_m,\; m>0$ corresponds to an $N$-fold
contraction; $N=1$ is the classical operator, $N=2$ is
the first quantum correction etc.. Explicitely,
\ba
&&\nspace
 Q_{\lb} = Q_{\lb}^{(0)} + \beta^2 Q_{\lb}^{(1)}
+ \ldots \nonum
&& \;\;Q_{\lb}^{(0)} = T_{\lb} \sum_{m\geq 1}
S_{m-1}[\lb\cdot x_n] \frac{1}{m} \d_m^{(\lb)}\;,\nonum
&& \;\; Q_{\lb}^{(1)} = T_{\lb} \sum_{m\geq 2}
S_{m-1}[\lb\cdot x_n] \left( \sum_{k\neq l,k+l=m}
\frac{1}{kl} \d_k^{(\lb)}\d_l^{(\lb)}
+\frac{1}{2}\sum_{2k =m}\frac{1}{k^2} (\d_k^{(\lb)})^2
\right)\;,\nonum
&& \;\; \mbox{etc.}\;,
\ea
where $\d^{(\lb)}_m =\sum_a\lb^a\dd{x_m^a}$.
In particular, the
operators $Q_i := Q_{-\alpha_i},\; i=1,\ldots,r$ associated
with minus the simple roots of $g$ are called screening
operators. In this case we set $\pi_i=\pi_{-\alpha_i},\;
x_n^i:=-\alpha_i\cdot x_n,\; \d_n^{(i)}:=
\d_n^{(\alpha_i)} =-\sum_{j=1}^r(\alpha_i,\alpha_j)\dd{x_n^j}$.
The Fock space representations $\pi_{\lb}$ can be given
the structure of a vertex operator algebra (VOA) or meromorphic
conformal field theory by augmenting a Virasoro
element and a realization of the derivative operator
\cite{mCFT}. The Virasoro element is given by
\bas
L_{-2} \is\frac{1}{2} x_1\cdot x_1 +
2(\beta^2\rho-\rho^{\vee})\cdot x_2\;,
\eas
where $\rho,\;\rho^{\vee}$ are the Weyl vector of $g$
and its dual. The derivative operator is realized
$\beta$-independently as
\bas
L_{-1} \is\sum_{j=1}^r\sum_{m>0} (m+1)x_{m+1}^j \d_m^{(j)}
- x_1^j x_0^j\;,
\eas
where $x_0^j$ is defined by $[x_0^j,\, T_{\lb}] =
(\lambda_j, \lb) T_{\lb}$ and $\lb_j$ is the
$j$-th fundamental weight of $g$. From the recursion relation
for the Schur polynomials one checks $[L_{-1},\, Q_i]=0,\;
1\leq i\leq r$. In the VOA one has a vector-operator
correspondence, which associates a unique sequence of
linear bounded operators $P_n,\;n\in \Z$
to any vector $P$ in its (pre-) Hilbert space.
The sequence is defined in terms of formal series
(`fields') $P(z)=\sum_n P_n z^{-n-deg(P)}$ on which two
types of composition maps (operator product expansion
and normal ordering) are defined. For the Fock spaces
$\pi_{\lb}$ the correspondence takes the form
\bas
\pi_0 \ni P = (m_1!\,x_{m_1}^{a_1})
      \ldots (m_n!\,x_{m_n}^{a_n})
& \rra & P(z)=:i\d^{m_1}\phi^{a_1}(z)
           \ldots i\d^{m_n}\phi^{a_n}(z):\;,
\nonum
\pi_{\lb}\ni P\otimes v_{\lb} &\rra &
:P(z)\;e^{i\lb\cdot\phi(z)}:
\eas
On the r.h.s. $:\;\;:$ denotes normal ordering in the
oscillators $\{a_n^a\}$ and $\phi^a(z)$ are
(free massless) fields with operator product
$\phi^a(z) \phi^b(w) = -\beta^2\delta^{ab}\ln(z-w)$.
In particular, we denote by $\cF_{\lb}$ the
space of linear bounded operators $\pi_{\lb'}\ra
\pi_{\lb+\lb'}$ which are obtained as formal
residues $\oint P e^{i\lb\cdot\phi}$ of the fields
corresponding to elements
$P\otimes v_{\lb}$ in $\pi_{\lb}$. Then
$\oint :\pi_{\lb}/L_{-1}\pi_{\lb} \ra \cF_{\lb}$
provides an isomorphism of linear spaces.

\bigskip
{\em 2.2.} Next we recall some facts from the cohomological
construction of $W$-algebras. Consider first the case where
$\beta^2$ takes numerical values. For all positive irrational
values of $\beta^2$, both the $W$-algebra itself and its
irreducible highest weight representations can be described
as the only non-vanishing cohomology class of a Fock space
resolution. The resolution is of a form similar to the
Bernstein-Gelfand-Gelfand (BGG) resolution
known for the irreducible $g$-modules. The Fock spaces
involved are parametrized by elements $\lb\in h^*$ lying in
an orbit of the Weyl group $W$ of $g$. Let $\alpha_i^{\vee}
\in h,\;1\leq i\leq r$ be the simple co-roots and
$\bra\,,\,\ket$  the duality pairing. The Weyl group acts
by $r_i*\lb = r_i(\lb +\rho) -\rho =\lb - \bra \lb
+\rho,\alpha_i^{\vee}\ket\alpha_i$ on $h^*$, where $r_i$
are the fundamental reflections. In the present context it
suffices to consider the Weyl group orbits of dominant
integral weights $\Lambda\in P_+$ of $g$. The resolution is
of the form\cite{MN2,FF2}
$$
0\ra \pi_{\Lambda}(0) \stackrel{d_{\Lambda}(0)}{\ra}
        \pi_{\Lambda}(1) \stackrel{d_{\Lambda}(1)}{\ra}
\ldots \stackrel{d_{\Lambda}(|\Delta_+|-1)}{\ra}
\pi_{\Lambda}(|\Delta_+|)\ra 0\;,
$$
where $\pi_{\Lambda}(k)=\bigoplus_{\{w\in W|l(w)=k\}}
\pi_{w*\Lambda}$ and $\Delta_+$ is the set of positive
roots. The differentials
$d_{\Lambda}(k),\;0\leq k\leq |\Delta_+|-1$ are
certain composite operators built from the $Q_i$'s.
We will denote the composition map by $[\;\;]$.
The lowest differential $d_{\Lambda}(0)$ is the direct sum of
$[Q_i]^{l_i+1}:\pi_{\Lambda}\ra \pi_{r_i*\Lambda}$, where
$l_i=(\alpha_i,\Lambda),\;1\leq i\leq r$. These are linear
operators, so that their kernels are graded subspaces of
$\pi_{\Lambda}$. For $\Lambda \in P_+$ consider
\be
\cH_{\Lambda}(g) = \bigcap_{i=1}^r
\Ker\left([Q_i]^{l_i+1}:\pi_{\Lambda}\rra
\pi_{r_i*\Lambda}\right)\;.
\ee
This space appears as the lowest, and only non-vanishing
cohomology class of the resolution, $\cH_{\Lambda}(g)=
\Ker d_{\Lambda}(0)$. Moreover it can be
shown to form an irreducible $W(g)$-module. Applying
the Euler-Poincare' principle to the resolution one
finds for the character
\be
ch_{\Lambda}(g) =\Tr_{\cH_{\Lambda}(g)}[t^{L_0}]=
\frac{t^{h_{\Lambda}}\prod_{\alpha\in\Delta_+}
      (1-t^{(\Lambda+\rho,\alpha)})}%
     {\prod_{n\geq 1}(1-t^n)}\;,
\ee
where $h_{\Lambda}=\frac{\beta^2}{2}(\Lambda,\Lambda+
2(\beta^2\rho -\rho^{\vee}))$ is defined through
$L_0v_{\Lambda}=h_{\Lambda}v_{\Lambda}$.

The relations (3) and (4) hold for all positive irrational
values of $\beta^2$. In the present context we wish to
treat $\beta$ as a formal variable not taking numerical
values. In this case one encounters the problem that for
any $P\in \bigcap_{i=1}^r\Ker[Q_i]^{l_i+1}$ also
$\beta^{2n} P,\;n>0$ are elements of
$\bigcap_{i=1}^r\Ker[Q_i]^{l_i+1}$ and as elements of the
$\C[\beta^2]$-module $\pi_{\Lambda}$ they are formally
inequivalent. There are several ways to avoid this
overcounting; the one which is technically most useful
refers to the classical limit of the above construction.
Define $\cH_{\Lambda}^{(0)}(g)\subset
\pi_{\Lambda}^{(0)}$ by
\be
\cH_{\Lambda}^{(0)}(g) = \bigcap_{i=1}^r
\Ker\left([Q_i^{(0)}]^{l_i+1}:\pi_{\Lambda}^{(0)}\rra
\pi_{r_i*\Lambda}^{(0)}\right)\;.
\ee
Also the space $\cH_{\Lambda}^{(0)}(g)$ appears as the only
non-vanishing cohomology class of a BGG-type resolution,
which it can be obtained as the $\beta\ra 0$ limit
of the `quantum' BGG resolution defining $\cH_{\Lambda}(g)$
in (3). In particular, the composition map for the operators
$Q_i^{(0)}$ (for simplicity also denoted by $[\;\;]$) arises
as the $\beta\ra 0$ limit of the composition map for
the $Q_i$'s. Notice that all the quantities in (5) are
$\beta$-independent and no ambiguity arises. Set now
$\widetilde{\pi}_{\Lambda}=
{\cal H}_{\Lambda}^{(0)}\oplus \beta^2 \pi_{\Lambda}$,
where $\pi_{\Lambda}=\pi_{\Lambda}^{(0)}\otimes\C[\beta^2]$.
In particular $\widetilde{\pi}_{\Lambda}$ is no longer a
module over $\C[\beta^2]$ and no overcounting occurs if
one considers
\be
{\cal H}_{\Lambda}(g) =\bigcap_{i=1}^r
\mbox{Ker}\left([Q_i]^{l_i+1}:\widetilde{\pi}_{\Lambda}
\longrightarrow \widetilde{\pi}_{r_i*\Lambda}\right)
\ee
(where $\widetilde{\pi}_{r_i*\Lambda}$ could also be
replaced with $\beta^2\pi_{r_i*\Lambda})$. Generally one
can define the  BGG-type resolution as before with the
Fock spaces $\pi_{w*\Lambda}$ replaced by
$\widetilde{\pi}_{w*\Lambda}$. Again the space
$\cH_{\Lambda}(g)$ appears as the only non-vanishing
cohomology class of the complex and is an irreducible
$W(g)$-module. In particular the character formula (4)
is preserved, but now with $\beta^2$ considered as a
formal variable not taking numerical values. The relation
between $\cH_{\Lambda}(g)\subset \widetilde{\pi}_{\Lambda}$
and its classical counterpart $\cH^{(0)}_{\Lambda}(g)
\subset \pi_{\Lambda}^{(0)}$ can be summarized as follows.
For $\Lambda \in P_+$:
\smallskip

\begin{itemize}
\item[$i.$] The restriction $C_{\Lambda}$ of $C$ to
$\cH_{\Lambda}(g)$ defines as isomorphism of graded linear
spaces $C_{\Lambda}:\cH_{\Lambda}(g)\rra
\cH_{\Lambda}^{(0)}(g)$.
\item[$ii.$] Let $P^{(0)}\in\cH_{\Lambda}^{(0)}$ be of degree
$N$ and power $m$. Then $P=C_{\Lambda}^{-1}(P^{(0)})$ is of
the form
\be
P = P^{(0)} + \beta^2 P^{(1)} +\ldots +
\beta^{2(m-1)} P^{(m-1)}\;,
\ee
where $P^{(i)}$ is of power $p\leq m-i$ in
$\{ x_n^j \}$. The term of leading power in $P^{(0)}$
(and hence $P$) is a a Weyl invariant polynomial,
where the Weyl group acts by
$w x_n^i= (w\alpha_i)\cdot x_n$ and
$w (x_n^{i_1}\ldots x_n^{i_k}) =
(w x_n^{i_1}\ldots w x_n^{i_k}),\;wv_{\Lambda}
=v_{\Lambda}$ on $\pi_{\Lambda}^{(0)}$.
\item[$iii.$] If $g$ is simply laced the elements
$P\in\cH_{\Lambda}(g)$ are of the form
\be
P=\sum_i c_i(\beta^2)\,X_i\;,
\ee
where $X_i\in \pi_{\Lambda}^{(0)}$ are monomials in
$\{x_n^i\}$. The coefficients $c_i(\beta^2)$ are polynomials
in $\beta^2$ of the form $c_i(\beta^2)=c_i +o(\beta^2)$,
with $c_i\neq 0$.
\end{itemize}
\smallskip

\noindent
Let us verify the claims $i.$--$iii.$ consecutively.
$i.$ Being the lowest cohomology
class of a BGG-type resolution, the character
$ch_{\Lambda}^{(0)}(g)$ of $\cH_{\Lambda}^{(0)}(g)$
can again be obtained from the Euler-Poincare' principle.
The result just differs by a factor from (4):
$ch_{\Lambda}(g)=t^{h_{\Lambda}}ch_{\Lambda}^{(0)}(g)$.
This implies that $\cH_{\Lambda}(g)$ and
$\cH_{\Lambda}^{(0)}(g)$ are isomorphic as graded
linear spaces. It remains to show that the isomorphism
is given by the mapping $C_{\Lambda}$. Since by construction
$C_{\Lambda}\cH_{\Lambda}(g)\subset\cH_{\Lambda}^{(0)}(g)$
it suffices to show that $C_{\Lambda}$ is injective i.e.
that some $P\in\cH_{\Lambda}(g)$ is uniquely determined by
its classical part $C_{\Lambda}(P)$. From the form of $Q_i$
in eqn.~(2) it is clear that $P\in\cH_{\Lambda}(g)$
has a series expansion of the form (7) with $P^{(0)}=
C_{\Lambda}(P)$. The uniqueness of this expansion follows
from the fact that $P^{(0)}$ is
assumed to be $\beta$-independent i.e. ad-hoc
$\beta$-dependent linear combinations of elements
in $\bigcap_{i=1}^r \Ker Q_i^{(0)}$ are excluded by
defining it to be a subspace of $\pi_0^{(0)}$:
Suppose two states $P$ and $\widetilde{P}$ in $\cH(g)$
to be given with the same classical part $P^{(0)}=C(P)
=C(\widetilde{P})\in \cH^{(0)}_{\Lambda}(g)$.
Then their first quantum corrections $P^{(1)}$ and
$\widetilde{P}^{(1)}$ have to differ by an element of
$\bigcap_{i=1}^r \Ker Q_i^{(0)}$. If non-zero, this
element would give rise to a $\beta$-dependent
contribution to the classical part of at least one
of the states $P$ or $\widetilde{P}$ -- in conflict with
the assumption. Hence $P^{(1)} = \widetilde{P}^{(1)}$.
By induction one concludes $P^{(k)}=
\widetilde{P}^{(k)},\;1\leq k\leq N-1$. Thus
$P=\widetilde{P}$, and $C_{\Lambda}$ is injective.

$ii.$ For the first part of $ii.$ it remains to show that
the corrections $P^{(i)}$ in (7) are of power less or equal
to $m-i$ in $\{x_m^i\}$. To check this, decompose
$\cH_{\Lambda}:=\cH_{\Lambda}(g)$ into the space of
non-derivative states
$\hH_{\Lambda}=\cH_{\Lambda}/L_{-1}\cH_{\Lambda}$ and
its orthogonal complement. If $P_k,\;k\in I$ is a basis
of $\hH_{\Lambda}$, the states $L_{-1}^n P_k,\;k\in I$
form a basis of the orthogonal complement. Thus, it
suffices to show the statement for $\hH_{\Lambda}$. Let
$(\widehat{\pi}_{\lb})_{N,p}$ be the subspace of
$\pi_{\lb}/L_{-1}\pi_{\lb}$ of degree $N$ and power $p$.
One then checks that $Q_{\lb}^{(k)}$ maps
$(\widehat{\pi}_{\lb})_{N,m}$ to
$(\widehat{\pi}_{\lb})_{N-1,m-k-1}$ (see the proof of
eqn.~(17) below). This implies (7) for $\hH_N$.

To verify the second part of the claim $ii.$ it is
convenient to make use of the fact that there is a
bijection from a space $\Omega_{\Lambda}$ of $g$-singlets
in level 1 $\gh$-modules to $\cH_{\Lambda}$.
On $\Omega_{\Lambda}$ one
can introduce a double graduation (power and degree) s.t.
the bijection preserves both gradings. Explicitely the
image of some $P\in \Omega_{\Lambda}$ is obtained by
projecting its $\phi\beta\gamma$ (Wakimoto-type) free
field realization onto the $\beta\gamma$-independent
part\cite{MN2}. These images are elements of
$\cH_{\Lambda}$, and by construction the term of leading
power is a Weyl invariant polynomial.

$iii.$ Let $g^{\vee}$ be the Lie algebra dual to $g$ (i.e.
the Cartan matrix of $g^{\vee}$ is the transpose of that
of $g$) and let $\beta$ momentarily assume numerical
values. Denote by $\widetilde{Q}_i$ and
$\widetilde{Q}_i^{\vee}$ the screening
operators obtained by replacing $x_n^i=-\alpha_i\cdot x_n$
with $\tilde{x}_n^i=-\beta\alpha_i\cdot x_n$ and
$(\tilde{x}_n^i)^{\vee}=\beta^{-1}\alpha_i^{\vee}\cdot x_n$,
respectively (identifying $h$ with $h^*$). For $\beta^2$
irrational there exists a class of irreducible $W(g)$-modules
$\widetilde{\cH}_{\Lambda_+,\Lambda_-}$ labeled by
$\Lambda_+\in P_+,\;\Lambda_-\in P_+^{\vee}$. These can
be described either as
$\bigcap_{i=1}^r [\widetilde{Q}_i]^{(\alpha_i,\Lambda_+)+1}$
or as
$\bigcap_{i=1}^r
[\widetilde{Q}_i^{\vee}]^{(\alpha_i^{\vee},\Lambda_-)+1}$
on $\pi_{\lb},\;\lb=\beta\Lambda_+-\beta^{-1}\Lambda_-$.
In other words, there exists an isomorphism of
$W(g)$-modules
\ba \nspace &&
\widetilde{\cH}_{\Lambda_+,\Lambda_-}\Big|_{\beta}=
\widetilde{\cH}_{\Lambda_-^{\vee},\Lambda_+^{\vee}}
\Big|_{-1/\beta}\;,
\ea
where $\Lambda^{\vee}$ is defined by $(\Lambda,\alpha_i)=
(\Lambda^{\vee},\alpha_i^{\vee})$. For $\Lambda_+
=\Lambda_-=0$ this was shown in \cite{FF2} and by
analysing the Kac determinant it can be extended to
$\Lambda_+\in P_+,\;\Lambda_-\in P_+^{\vee}$: It is known
that the singular vectors in the Verma module $M(\lb)$ can
in the free field realization be constructed from
$W(g)$-intertwiners $\pi_{\lb'}\ra\pi_{\lb}$. For
weights of the form $\lb = \beta\Lambda_+
-\beta^{-1}\Lambda_-$, where $\Lambda_+\in P_+,\;
\Lambda_-\in P_+^{\vee}$ the Verma module contains a
singular vector at degree $(\Lambda_+ +\rho,\alpha_i^{\vee})
(\Lambda_- +\rho^{\vee},\alpha_i)$ for each
$1\leq i\leq r$. The degree of these (and hence all
other) singular vectors is left unchanged by the duality
transformation $g\ra g^{\vee},\;\beta \ra -\beta^{-1}$
if $\Lambda_{\pm}\ra \Lambda_{\mp}^{\vee}$.

The case at hand corresponds to $(\Lambda_+,\Lambda_-)=
(\Lambda,0)$. Applied to simply-laced Lie algebras $g$ the
duality (9) implies $\widetilde{\cH}_{\Lambda,0}
\big|_{\beta}=\widetilde{\cH}_{0,\Lambda}\big|_{-1/\beta}$.
If $\beta^2$ is irrational also $1/\beta^2$ is irrational,
so that $ch\widetilde{\cH}_{\Lambda,0}=
ch\widetilde{\cH}_{0,\Lambda}$ (even for the same $\beta$).
Thus, in fact $\widetilde{\cH}_{\Lambda,0}=
\widetilde{\cH}_{0,\Lambda}$. If we now restore $\beta$ to
be a formal variable this forces the elements
$\widetilde{P}[\{\tilde{x}_n^i\}]$ of
$\widetilde{\cH}_{\Lambda,0}=\widetilde{\cH}_{0,\Lambda}$
to be invariant under
the substitution $\beta\ra -1/\beta$ i.e. to be functions of
$\tilde{B}=\beta -1/\beta$ only. If $\widetilde{P}$ is of
leading power $m$ in $\{\tilde{x}_n^i\}$, one returns to
the original normalizations via
$P[\{x_n^i\}]=\beta^m\widetilde{P}[\{\tilde{x}_n^i/\beta\}]$.
This implies that elements $P$ of $\cH_{\Lambda}(g)$, rather
then being polynomials in $\beta^2$, in fact are polynomials
in $B=1-\beta^2$. When rewritten in the form (7) the elements
$P\in\cH_{\Lambda}(g)$ are seen to be of the form (8).
\medskip

\noindent{\em Remark:}
Of particular interest is the space
$\cH(g):=\cH_0(g)$, which
encodes information about the operator product expansion
of the $W(g)$-algebra. By taking $\cH(g)$ to be
the pre-Hilbert space of a VOA one can show that the
quantum $W(g)$-algebra exists if and only if $ch_0(g)$
coincides with the character of the free polynomial
algebra generated by elements $W^i_{n_i},\; 1\leq i\leq r,
\;n_i\geq -(e_i+1)$, where $e_i$ are the exponents of $g$.
Explicitely, the condition reads
\be
ch_0(g) \stackrel{\textstyle !}{=}
\prod_{1\leq i\leq r,\;n_i>e_i}
(1-q^{n_i})^{-1}\;.
\ee
In order to prove the existence of a quantum $W(g)$-algebra
one thus has to establish (10) from the defining relation (3).
In \cite{MN2,FF2} this was achieved by extending
$\cH_{\Lambda}(g)$ to a BGG-type resolution.
It then follows that $\cH(g)$ contains a set
of fundamental vectors $W^i_{-e_i-1},\; 1\leq i\leq r$
from which one can build states of the form
\ba
\nspace \!\!\!&& W^{i_1}_{-m_1-1-e_{i_1} }\ldots
W^{i_n}_{-m_n-1-e_{i_n} }\otimes v_0,\;\;\;
i_j\geq i_{j+1},\; m_j \geq 0\;,\;\;
m_j\geq m_{j+1}\;\; \mbox{if}\;\; i_j =i_{j+1}.
\ea
The validity of the condition (10) implies in particular
that these states are linearly independent and exhaust all
solutions to $Q_i\, P =0,\; 1\leq i\leq r$ at a
given degree.
\medskip

Consider now the spaces $\cH_{\Lambda}(g)$ modulo
$L_{-1}$-exact pieces. Set $\hH_{\Lambda}:=
\hH_{\Lambda}(g):=
\newline  \cH_{\Lambda}(g)/L_{-1}\cH_{\Lambda}(g)$, which
inherits the $\deg \, x_n^i = n$ grading from $\cH_{\Lambda}$.
Let $(\hH_{\Lambda})_N$ denote the subspace of degree $N$.
As before the classical counterparts will carry a
superscript `$(0)$'. Let $\Ch_{\Lambda}$ denote the restriction
of $C$ to $\hH_{\Lambda}$. From the previous results
it follows that this is a bijection of graded vector spaces
\be
\Ch_{\Lambda} :\hH_{\Lambda}\rra \hH_{\Lambda}^{(0)}\;,
\bspace \Lambda \in P_+\;.
\ee
%%%%%%%%%%%%%%%%%%%%%%%%%%%%%%%%%%%%%%%%%%%%%%%%%%%%%%%%%%

\newsection{Existence of infinite abelian subalgebras}
{\em 3.1.} The commutation relations of generic elements
of $W(g)$ can be written in an $su(1,1)$-covariant form,
in which all the specific information is encoded in
the structure constants (the latter being polynomial
or algebraic functions of the central charge). Given
this form of the commutation relations one can
investigate the conditions under which $W(g)$ possesses
abelian subalgebras. It turns out that the existence of
an abelian subalgebra amounts to the existence of a
preferred basis in which a certain subset of the
structure constants vanishes\cite{MN3}. The existence
of an infinite dimensional abelian subalgebra in $W(g)$
therefore is an important structural result, somewhat
analogous to the existence of a Cartan subalgebra for
ordinary Lie algebras.

A first understanding of this result can be obtained
from the classical theory. The projection $C$ onto
the $\beta$-independent part defines a contraction
of the $W(g)$-algebra, which coincides with the classical
$W$-algebra $W^{(0)}(g)$. These classical algebras
$W^{(0)}(g)$ are known to appear as symmetry algebras
in various classical integrable systems (mKdV, KdV,
affine Toda). Their conserved charges can be expressed
in terms of the generators of the classical $W$-algebra
and generate an infinite dimensional abelian subalgebra
thereof. This result can be derived from any of the
various approaches to classical integrable systems.
The formulation which is most useful for the transition
to the quantum theory is the following.

Let $\alpha_0$ denote the additional simple root
associated with the Dynkin diagram of an affine
Lie algebra $\gh$ and set $\cF:=\cF_{\alpha=0},\;Q_0
:=Q_{-\alpha_0}$,
$\cF_i:= \cF_{-\alpha_i},\;0\leq i\leq r$. Since $\alpha_0$
enters only through inner products with elements of $h$ we
can identify $\alpha_0$ with its projection $-\theta$ on $h$
(For $\gh=A_{2r}^{(2)}$ this differs by a factor 2 from the
standard definition). Note that $\theta\in P_+$, so
that the results of section 2 apply to $\cH_{\theta}$ and
$\hH_{\theta}$. The classical counterparts of these objects
will again carry a superscript `$(0)$'.
On the classical level one can show that
\bas
&& I^{(0)}(\gh) := \bigcap_{i=0}^r \Ker
\left( Q_i^{(0)}:\cF^{(0)} \rra \cF_i^{(0)} \right)
\eas
defines an abelian Poisson-subalgebra of the classical
$W$-algebra $W^{(0)}(g)$ which is linearly generated by
elements $I^{(0)}_N=\oint J_N^{(0)}$, with $J_N^{(0)}$
of degree $N,\;N-1\in E$. The elements
$J^{(0)}_N$ are called the classical conserved
densities at degree $N$. By definition they are taken to be
$\beta$-independent. The index set $E=E(\gh)$ are the
exponents of $\gh$, an infinite subset of $\N$ uniquely
associated to $\gh$. For later use note the following
properties of the conserved densities.

\begin{itemize}
\item[--] Nonvanishing classical conserved densities
$J_N^{(0)}$ exist only for degree $N-1\in E$.
\item[--]
The number of linearly independent conserved densities
at degree $N$ equals the multiplicity $\#(N-1)$ of the
exponent.
\item[--] Their linear hull does not contain elements
of leading power less then $N$ in $\{x_n^i\}$.
\end{itemize}

This follows from the results of Drinfel'd and Sokolov%
\cite{DS}. The first two facts are stated explicitely
(remarks to prop. 6.6) and the last one can be
extracted from the proof of prop. 6.2. We note that
only algebras of type $D_{2n}^{(1)},\;n>1$ have exponents
of multiplicity greater than 1 (the exponents $2n-1
\bmod (4n-2)$ appear with multiplicity two). In all other
cases the conserved densities at some degree $N,\;N-1\in E$
are unique. The aim of this paper is to show that analogous
results hold in the quantum theory.

\bt{Theorem}
\bas
&& I(\gh) := \bigcap_{i=0}^r \Ker
\left( Q_i:\cF \rra \cF_i \right)
\eas
defines an abelian subalgebra of the quantum
$W(g)$-algebra. $I(\gh)$ is linearly generated by elements
$I_N=\oint J_N$, $N-1\in E$ (including multiplicities).
\et
\bigskip

{\em 3.2.} The problem consists in proving the existence
of the elements $J_N$. The remaining properties then follow
easily. One checks that $[Q_0,L_{-1}]=0$ i.e. $L_{-1}\in I(\gh)$.
If we anticipate that $Q_0^{(0)}$ maps $\hH_0^{(0)}$ to
$\hH_{\theta}^{(0)}$ (c.f. the paragraph following
eqn.~(13)) this means that $Q_0$ maps
$\widetilde{\pi}_0/L_{-1}\widetilde{\pi}_0$ to
$\widetilde{\pi}_{\theta}/L_{-1}\widetilde{\pi}_{\theta}$.
In view of (6) the intersection of the kernels
$\bigcap_{i=0}^r \Ker Q_i$
on $\widetilde{\pi}_0/L_{-1}\widetilde{\pi}_0$
defines a subspace of $\hH =\hH_0(g)$. All the elements of
$I(\gh)$ can thus be expressed in terms of $W$-generators
and since for $I_N,\;I_M\in I(\gh)$ also $[I_N,I_M]\in
I(\gh)$ one concludes that $I(\gh)$ defines a subalgebra
of $W(g)$.%
\footnote{It is only for this conclusion that the
construction via vertex operators i.e. the
condition (13) below is essential. If the elements
$\oint \Ch^{-1} J_N^{(0)}$ could independently be shown
to form a subalgebra, the result on the existence of an
infinite abelian subalgebra would follow already at this
point.}
\begin{samepage}
We wish to show that it is in fact an abelian
subalgebra. To see this, recall that the quantum conserved
densities $J_N$ are elements of $\hH_N$. One may thus
make use of the bijection (12). If for some $N-1\in E$
there exists a
quantum conserved density $J_N$ at all, its classical
part $\Ch J_N$ has to be a classical conserved density.
Conversely, given a classical conserved density
$J_N^{(0)}\in \hH_N^{(0)}$ there is only one possible
candidate for a quantum conserved density namely
$\Ch^{-1}(J_N^{(0)})$. In particular it follows  that the
quantum conserved density $J_N$ associated with $J_N^{(0)}$,
if any, must be of leading power $N$ in $\{x_n^i\}$. This
forces $I(\gh)$ to be abelian: If $J_N(z) =\sum_n (J_N)_n
z^{-n-N}$ denotes a conserved density of degree $N$, the
conserved charge is the mode of degree $-N+1$, i.e.
$I_N =\oint dz J_N(z) =(J_N)_{-N+1}$. In this situation
the general form of the commutator in a VOA reduces to
\cite{MN3}
\end{samepage}
\pagebreak
\bas\nspace
&& [(J_M)_{-M+1}\,,\,(J_N)_{-N+1}]=\sum_{K:M+N-K=1}
C_{MN}^{K}\;(P_K)_{M+N-2}\;,\sspace N-1,M-1\in E\;,
\eas
where $C_{MN}^K$ are the structure constants coupling to
the fields $P_K(z)$ in the VOA which have degrees $K$
satisfying $M+N-K=1$. Since $I(\gh)$ is a subalgebra of
$W(g)$ the r.h.s. must
again be an element of $I(\gh)$. In particular every
field $P_K(z)$ for which the structure constant $C_{MN}^K$
were nonvanishing would have to be of leading power
$M+N-1$ in $\d\phi^a$. But the terms of power $M+N-1$
on the r.h.s., if any, are the same in the classical and the
quantum case. In the classical case they are known to be
absent, leaving a r.h.s. which is of leading power
less than $M+N-1$ and which thus cannot be a conserved
charge. Hence the structure constant $C_{MN}^K$ has to vanish.
In particular this shows that the existence of an abelian
subalgebra in $W(g)$ is equivalent to the vanishing of
the set of structure constants $C_{MN}^K$ with $K$ running
over all fields in the VOA of degree $K=M+N-1$.
\medskip

To prove the theorem it therefore suffices to show that for
every classical conserved density $J^{(0)}_N$ the unique
candidate for its quantum counterpart $\Ch^{-1} J^{(0)}_N$
is annihilated by $Q_0$ (modulo $L_{-1}$-exact pieces) i.e.
\be
Q_0\,[\Ch^{-1}J_N^{(0)}]\stackrel{\mbox{!}}{=}
L_{-1}\left( S\otimes v_{\theta}\right)\;,
\ee
for some $S\in \pi_0$. The essential point in proving (13)
will be that the r.h.s.
in fact is an element of $\hH_{\theta}$ on which the mapping
$\Ch_{\theta}$ exists and is invertible. For this one has
to show that $Q_0$ is a well-defined mapping from $\cH$ to
$\cH_{\theta}\;(*)$. Since $Q_0$ commutes with $L_{-1}$ it
then follows that $Q_0$ induces a mapping between the
equivalence classes modulo total derivatives in $\cH$ and
$\cH_{\theta}$, respectively. In particular, these
results will hold for the restrictions to the classical
part.

To check $(*)$ recall that
both $\cH$ and $\cH_{\theta}$ are irreducible $W(g)$-%
modules. Thus in effect one has to construct a free field
realization of an intertwiner connecting both $W(g)$-%
modules. This can be done via the BGG-type resolution by
constructing a sequence of intertwiners $V_{ww'}^{(i)}:
\pi_{w*0}\ra\pi_{w'*\theta}$ connecting the Fock spaces
in both resolutions which are labeled by Weyl group
elements $w,w'$ of the same length. These intertwiners
have to satisfy a number of consistency conditions which
allows one to determine them explicitely\cite{BMP}.
The result is that the whole sequence of intertwiners
$V^{(i)}=\{V_{ww'},\;l(w)=l(w')=i\},\;
0\leq i\leq |\Delta_+|$ is already determined by $V^{(0)}$
and $V^{(1)}$. Explicitely the latter are given by
\bas
&& V^{(0)}=\{Q_0\}\;,\bspace
V^{(1)}=\{Q_{ij}=\delta_{ij}
[Q_0\,Q_j^{(\theta,\alpha_j)}]\}\;.
\eas
These intertwiners make the diagram in Fig.~1 commutative.
In particular it follows that elements in $\cH$ are
mapped into elements of $\cH_{\theta}$.
%%%%%%%%%%%%%%%%%%%%%%%%%%%%%%%%%%%%%%%%%%%%%%%%%%%%%%%%%
\begin{center}
\setlength{\unitlength}{1mm}
%mainpicture
\begin{picture}(160,75)
\thicklines
\put(0,70){Fig.1: Free field realization of the
$W(g)$-intertwiner $\cH \ra\cH_{\theta}$.}
%
%subpicture1
\put(25,50){\begin{picture}(150,60)\thicklines
\put(0,0){0}
\put(5,1){\vector(1,0){20}}
\put(30,0){$\widetilde{\pi}_{0}$}
\put(40,1){\vector(1,0){20}}
\put(65,0){$\widetilde{\pi}_{r_{i}*0}$}
\put(80,1){\vector(1,0){20}}
\put(105,0){$\ldots$}
\put(47,5){$[Q_{i}]$}
%vectors
\put(32,-5){\vector(0,-1){25}}
\put(70,-5){\vector(0,-1){25}}
\put(34,-18){$V^{(0)}$}
\put(72,-18){$V^{(1)}$}
\end{picture}}
%end subpicture1
%
%subpicture2
\put(25,10){\begin{picture}(150,60)\thicklines
\put(0,0){0}
\put(5,1){\vector(1,0){20}}
\put(30,0){$\widetilde{\pi}_{\theta}$}
\put(40,1){\vector(1,0){20}}
\put(65,0){$\widetilde{\pi}_{r_{j}*\theta}$}
\put(80,1){\vector(1,0){20}}
\put(105,0){$\ldots$}
\put(41,5){$[(Q_{j})^{(\theta,\alpha_{j})}]$}
\end{picture}}
%end subpicture2
%
\end{picture}
\end{center}
%%%%%%%%%%%%%%%%%%%%%%%%%%%%%%%%%%%%%%%%%%%%%%%%%%%%%%%%%%%%

{\em 3.3.} It remains to prove (13). First observe the
identity
\be
Q_0^{(0)}\, C = C_{\theta}\,Q_0 \;.
\ee
To check (14) let $P\in \cH$ be given. Then $Q_0P$ is an
element of $\cH_{\theta}\subset \widetilde{\pi}_{\theta}$
and for later use we write it in the form
$$
Q_0 \,P= R\otimes v_{\theta}+L_{-1}(S\otimes v_{\theta})\;,
$$
for some $R,\;S \in \pi_0$. Similarly for $P^{(0)}=CP$ one
has
$$
Q_0^{(0)} P^{(0)}= R^{(0)}\otimes v_{\theta}+
L_{-1}(S^{(0)}\otimes v_{\theta})\;,
$$
for some $R^{(0)},\;S^{(0)}\in \pi_0^{(0)}$. But since
$Q_0 P=Q_0^{(0)} P^{(0)} + o(\beta^2)$ and $C_{\theta}$
commutes with $L_{-1}$ it follows that $C_{\theta}
(R\otimes v_{\theta})=R^{(0)}\otimes v_{\theta}$,
$C_{\theta}(S\otimes v_{\theta})=S^{(0)}\otimes
v_{\theta}$ and thus $C_{\theta}Q_0P=Q_0^{(0)}CP$.
This holds for all $P\in \cH$ and implies (14). Moreover
since both $C_{\theta}$ and $C$ commute with $L_{-1}$
eqn.~(14)
projects to an identity $\Ch_{\theta}Q_0 =Q_0^{(0)}\Ch$
on $\hH$, expressing the commutativity of the diagram in
Fig.~2. But since both $\Ch_{\theta}$ and $\Ch$ are
invertible this means that $Q_0$ and $Q_0^{(0)}$ are related
by an equivalence transformation
\be
Q_0^{(0)} =\Ch_{\theta}\, Q_0\,\Ch^{-1}
\ee
and have the same rank (on all $\hH_N,\;N\geq 0$).
In particular eqn.~(15) implies that the kernel of
$Q_0^{(0)}$ on $\hH_N^{(0)}$ and the kernel of $Q_0$ on
$\hH_N$ have the same dimension and hence guarantees that
every classical conserved density can be deformed into a
quantum conserved density. Explicitely, if $J_N^{(0)}\in
\Ker Q_0^{(0)}$ on $\hH_N^{(0)}$ the unique candidate
$\Ch^{-1}J_N^{(0)}\in \hH_N$ for its quantum counterpart
will always solve (13): $Q_0\,\Ch^{-1}J_N^{(0)}=
\Ch^{-1}_{\theta} Q_0^{(0)} J_N^{(0)} =0\in
\hH_{\theta}$, which is what we wanted to show.
%%%%%%%%%%%%%%%%%%%%%%%%%%%%%%%%%%%%%%%%%%%%%%%%%%%%%%%%%
\begin{center}
\setlength{\unitlength}{1mm}
%mainpicture
\begin{picture}(160,75)
\thicklines
\put(0,73){Fig.2: Commutative diagram for
$\mbox{rank}\,Q_0^{(0)}=\mbox{rank}\,Q_0$.}
%
%subpicture1
\put(25,50){\begin{picture}(150,60)\thicklines
\put(30,0){$\hH$}
\put(40,1){\vector(1,0){20}}
\put(65,0){$\hH^{(0)}$}
\put(47,5){$\Ch$}
%vectors
\put(32,-5){\vector(0,-1){20}}
\put(67,-5){\vector(0,-1){20}}
\put(34,-16){$Q_0$}
\put(72,-16){$Q_0^{(0)}$}
\end{picture}}
%end subpicture1
%
%subpicture2
\put(25,15){\begin{picture}(150,60)\thicklines
\put(30,0){$\hH_{\theta}$}
\put(40,1){\vector(1,0){20}}
\put(65,0){$\hH_{\theta}^{(0)}$}
\put(47,5){$\Ch_{\theta}$}
\end{picture}}
%end subpicture2
%
\end{picture}
\end{center}
\vspace{-2cm}

%%%%%%%%%%%%%%%%%%%%%%%%%%%%%%%%%%%%%%%%%%%%%%%%%%%%%%%%%%%%
\newsection{Construction principle}
{\em 4.1.} In the classical theory there exist several
fairly efficient methods to compute the conserved
densities $J^{(0)}_N$ explicitely. So far none of these
methods could be transferred to the quantum theory. The
results (13), (15) show that the quantization of a
classical conserved density $J^{(0)}_N$ can in principle
be done by applying the mapping $\Ch^{-1}_N$ (i.e. the
restriction of $\Ch^{-1}$ to the required degree $N$).
Unfortunately no efficient (possibly recursive) method
to compute $\Ch_N$ is available. (Part of the problem
being that it is not a homomorphism w.r.t. normal ordering
i.e. $\Ch(:\!PQ\!:)\neq \Ch(:\!P\!:)\Ch(:\!Q\!:)$, in general.)
In this section we will describe a reasonably efficient
method to compute the conserved densities explicitely.
We will introduce two different types of basis ${\cal A}$ and
${\cal B}$ on $\hH_N$ which are characterized by certain
extremal properties. A basis of type ${\cal B}$ is adapted to
the operator $Q_0$ and gives an alternative understanding
of the result (15). In particular the conserved densities
are part of a basis ${\cal B}$. In order to construct these
basis vectors we exploit the interplay between a basis of
type ${\cal B}$ and some more easily accessible type of
basis ${\cal A}$, whose elements have a maximal staggering in
their leading powers.
\medskip

A basis $\cA = \bigcup_{m=2}^N \A{m}$ of $\hH_N$
is called {\em of type} ${\cal A}$ if
\begin{itemize}
\item[a)] $\A{m}$ contains the maximal number of
linearly independent elements $P\in \cH_N$ which
are exactly of leading power $m$ in $\{x_n^i\}$.
\item[b)] The linear hull $\LA{m}$ does not contain
elements of leading power less than $m$ in $\{x_n^i\}$.
\end{itemize}
\medskip

\noindent
{\em Remark:} Clearly a basis of type ${\cal A}$ in $\cH_N$
will induce a basis on $\hH_N$ with the same properties
$a)$ and $b)$. For simplicity we will keep the notation
$\cA=\bigcup_{m=2}^N\A{m}$ for such basis on $\hH_N$.
No confusion will be possible as in this subsection $\cA$
will always refer to a basis on $\cH_N$, while from
subsection {\em 4.2} on $\cA$ will always refer to a basis
on $\hH_N$.
\smallskip

Condition $b)$ implies that $\LA{m}\cap \sum_{p<m}
\LA{p} =\{0\}$. Clearly $\LA{m}$ has trivial
intersection also with $\sum_{p>m}\LA{p}$. This means
that $\cH_N$ is the direct sum of the linear hulls
\bas
\nspace &&\cH_N = \LA{N} \oplus \ldots \oplus \LA{2}\;,
\eas
thereby introducing a grading on $\cH_N$. Of course a
basis of this type is not unique -- each element of
$\A{m}$ can be modified by adding an element of
$\bigoplus_{p\leq m-1}\LA{p}$ -- but the dimensions
of the subspaces $\LA{m}$ are unique. The
calculation of the associated `power characters'
is an open problem. Notice also that $A_{N,1}$ satisfying
condition $a)$ of the definition would be empty, which
is why $m=1$ has been omitted in $\cA$. For $N>2$ also
$\A{2}$ is empty.

One can fix the ambiguity in the definition of a basis of
type ${\cal A}$ in the following way. Recall from eqn.~(7)
that for any $P\in \cH_N$ the term of leading
power is a Weyl invariant polynomial. From the character
formula of the spaces $\cH_N,\;\cH^{(0)}$ and $\Omega_0$
(defined in the proof of eqn.~(7)) it follows that $P$, its
classical part $P^{(0)}$, and the Weyl invarant polynomial
$P_0$ of leading power (c.f. eqn.~(18) below) all are in 1-1
correspondence
\be
P\stackrel{1-1}{\longleftrightarrow}P^{(0)}
\stackrel{1-1}{\longleftrightarrow}P_{0} \;.
\ee
In particular for $P\in \LA{m}$ the part $P_0$ is of
power $m$ and adding elements of $\bigoplus_{p\leq m-1}
\LA{p}$ to it will affect only Weyl invariant terms
of power less than $m$. Because of the bijection (16)
this means that one can always find a basis
${\cal A}'_N =\bigcup_{m=2}^N A'_{N,m}$ of type ${\cal A}$
s.t. for all $P\in L(A'_{N,m})$ the Weyl invariant
polynomial $P_0$ of power $m$ is in fact the only Weyl
invariant term in $P$.

The basis ${\cal A}'_N$ of $\cH_N$ can be
described explicitely. We claim that $A'_{N,m}$ is just given
by the set of states (11) of degree $N=\sum_i(m_i+e_i+1)$
and leading power $m=\sum_i(e_i+1)$. Denote the latter
set of states by ${A''}_{N,m}$. We wish to show
that ${A''}_{N,m}$ has all the properties
characterizing $A'_{N,m}$. Clearly ${A''}_{N,m}$
contains only linearly independent elements of $\cH_N$
which are exactly of leading power $m$. Further
$\bigcup_{m=2}^N {A''}_{N,m}$ is a basis of
$\cH_N$ so that no further linearly independent elements
with that property exist. Thus ${A''}_{N,m}$
satisfies condition $a)$ in the definition of a basis of
type ${\cal A}$. To check condition $b)$ one makes use of the
bijections (16). For the Weyl invariant polynomials $P_0$
condition $b)$ is manifestly satisfied and thus holds also
for the states $P\in {A''}_{N,m}$ associated with
them. Hence the basis
$\bigcup_{m=2}^N {A''}_{N,m}$ is of type ${\cal A}$.
It remains to check that the only Weyl invariant terms in
some $P\in {A''}_{N,m}$ are those of power $m$.
But this follows because in the generators $W^i_{-m},\;
m\geq e_i+1$ only the term of power $e_i+1$ is Weyl
invariant. We conclude that the basis $\bigcup_{m=2}^N
A'_{N,m}$ and $\bigcup_{m=2}^N {A''}_{N,m}$ coincide.

In particular, this yields an explicit construction for
a basis of type ${\cal A}$ in $\cH_N$. As remarked before
such a basis will induce a basis on $\hH_N$ with the
same properties $a)$ and $b)$. From {\em now on} only basis
of type ${\cal A}$ on $\hH_N$ will be needed and for simplicity
we will keep the notation $\cA=\bigcup_{m=2}^N\A{m}$
for such basis.
\bigskip

{\em 4.2.} The basis $\cA$ on $\hH_N$
are defined without reference to
the operator $Q_0$. Next we introduce a type of basis $\cB$
adapted to the operator $Q_0$. In particular the conserved
densities are part of the basis. In order to define this
type of basis some preparations are needed. Let
$(\hH_{\Lambda})_{N,m}$ denote the subspace of
$\hH_{\Lambda}(g)$ containing the elements of degree $N$
and leading power $m$. Recall the shorthand $\hH=\hH_0(g)$.
We claim that
\ba
&& Q_0:\hH_{N,m} \rra
(\hH_{\theta})_{N-1,m-2}
\ea
is well defined. (This holds only on the equivalence
classes modulo total derivatives. When acting on
$\cH_N$ the powers get mixed in a complicated fashion.)
\medskip

The fact that $Q_0$ maps $\hH$ to
$\hH_{\theta}$ has already been seen in section 3.2.
It remains to show that $Q_0$ reduces the degree and the
power by 1 and 2 units, respectively. We first show that
for generic elements $\lb\in h^*$ the operator $Q_{\lb}$
acts as
\ba
&& Q_{\lb}:(\widehat{\pi})_{N,m}\rra
\bigoplus_{p\leq m-1}
\left( \widehat{\pi}_{\lb} \right)_{N-1,p}\;,
\ea
where $(\widehat{\pi}_{\lb})_{N,p}$ is the subspace
of $\pi_{\lb}/L_{-1}\pi_{\lb}$ of degree $N$ and
power $p$. For the particular choice $\lb =\theta$ we
then show that when acting on $\hH_{N,m}$ the value
$m-1$ on the r.h.s. can in fact be replaced by $m-2$.

To check (18) it is convenient to make use of the
vector-operator correspondence in the VOA. The action of
the operator $Q_{\lb}$ on differential polynomials
in the fields $i\d^m\phi^a(z)$ associated
with the states $m!\,x_m^a$ is then induced by the
operator product expansion with the normal ordered
exponential $\beta^{-2}:\exp(i\lb\cdot\phi(w)):$.
Consider first a single contraction contribution:
\bas
\beta^{-2}\,:i\d_z^{(n)}\phi^a\, P(z):\;
:e^{i\lb\cdot\phi(w)}:& \sim &
\frac{\lb^a P(z)}{(z-w)^n} \;
:e^{i\lb\cdot\phi(w)}: +\; (I) \nonum
                          & \sim &
\frac{\lb^a \d_w^{(n-1)} P(w)}{z-w} \;
:e^{i\lb\cdot\phi(w)}: +\;(I)\; +\;(II)\;.
\eas
Here $(I)$ denotes terms coming from the contraction
with $P(z)$ and $(II)$ denotes terms of the form
$\frac{(\cdots)(w)}{(z-w)^n},\; n\geq 2$. Clearly, upon
integration $\oint_{z} dw$ the terms of
type $(II)$ give rise to total derivative terms
and the remainder has both the
power and the degree reduced by one unit. By induction
on the power of $P(z)$ one finds $Q_{\lb}^{(0)}:
(\widehat{\pi})_{N,m} \ra (\widehat{\pi}_{\lb})_{N-1,m-1}$.
for the single contractions. Similarly one checks for
the higher contractions $Q_{\lb}^{(k)}:
(\widehat{\pi})_{N,m} \ra (\widehat{\pi}_{\lb})_{N-1,m-k-1}$.
In particular this implies (17).

Now specialize to $\lb =\theta$. Recall from eqn.~(7)
that for $P\in \cH$ the term of leading power is a Weyl
invariant polynomial. For some $P\in \cH_N$ of leading
power $m$ this means
\be
P = P_0 +P_1(\beta^2)\;,
\ee
where $P_0\in \pi_0^{(0)}$ is a Weyl invariant polynomial
of power $m$ in $\{x_n^i\}$ and $P_1(\beta^2)\in \pi_0$
is of power less or equal to $m-1$.
Consider now the action of $Q_0$ on some $P\in \hH_N$. One
can consider $P$ also as an element of $\cH_N$ with the
total derivative term fixed by the condition $L_1 P=0$.
Then $Q_0 P =Q_0 P_0 + Q_0 P_1\;(*)$. But on a
Weyl invariant polynomial the action of
$T_{\theta}^{-1}Q_0$ and $T_{-\alpha_i}^{-1}Q_i,\;
1\leq i\leq r$ coincide, so that $Q_0 P_0 =
T_{\theta}T_{-\alpha_i}^{-1}Q_i P_0 =
T_{\theta}T_{-\alpha_i}^{-1}Q_i (P -P_1) =
-T_{\theta}T_{-\alpha_i}^{-1}Q_i P_1$.
Inserting into $(*)$ this becomes
$Q_0 P = (Q_0 - T_{\theta}T_{-\alpha_i}^{-1}Q_i) P_1$.
Since $P_1$ is an element of $\LA{m-1}$, the r.h.s.
by (18) lies in $\bigoplus_{p\leq m-2}
\left(\widehat{\pi}_{\theta}\right)_{N-1,p}$.
This completes the proof of eqn.~(17).
\medskip

We shall need also the classical counterpart of eqn.~(17)
stating that $Q_0^{(0)}$ maps $\hH^{(0)}_{N,m}$ to
$(\hH^{(0)}_{\theta})_{N-1,m-2}$.
Indeed eqn.~(17) does hold also classically. To check
this note that equation (18) in particular holds for the
single contraction (classical) contributions. In addition
equation (19) is preserved with $P_1(\beta^2)$ replaced by
$P^{(0)}_1=\Ch P_1$, so that also the second part of the
proof goes through.
Eqn.~(17) and its classical counterpart can now be used to
define a special class of basis on $\hH_N$ and
$\hH^{(0)}_N$, respectively.
\medskip

A basis $\cB$ of $\hH_N$ is called of {\em type} ${\cal B}$ or
{\em extremal w.r.t.} $Q_0$ if $\cB = \bigcup_{m=2}^N \B{m}$,
$\;\;\;\;\B{m} \subset \hH_N$ and
\vspace{-2mm}

\begin{itemize}
\item[a)] $\B{m}$ contains the maximal number of
linearly independent elements $P\in \hH_N$  for which
$Q_0\,P$ is of the form
\be
Q_0\, P = R\otimes v_{\theta}  +
L_{-1}(S\otimes v_{\theta})\;,
\ee
where $R\in \pi_0$ is of exactly leading power
$m-2$ in $\{x_n^i\}$.
\item[b)] The linear hull $\LB{m}$ does not contain elements
satisfying (20) with $R$ of power less than $m-2$.
\end{itemize}
\medskip

{}From condition $b)$ one infers that $\LB{m}\cap
\sum_{p\neq m}\LB{p} =\{0\}$, so that
$\hH_N$ is the direct sum of the linear hulls
\bas
&& \hH_N = \LB{N} \oplus \ldots \oplus \LB{2}\;.
\eas
Of course a basis of this type is not unique
-- each element of $\B{m}$ can be modified by
adding an element of $\bigoplus_{p\leq m-1}\LB{p}$ --
but the dimensions of the subspaces $\LB{m}$ are unique.
The space $\LB{2}$ is uniquely defined however
and is just the space of quantum conserved densities at
degree $N$.

Similarly one can define a basis of $\hH^{(0)}_N$ which
is extremal w.r.t. $Q_0^{(0)}$.
Let ${\cal B}_N^{(0)} =\bigcup_{m=2}^N
\Bo{m}$ denote such a basis. One would expect that
$\Ch^{-1}\Bo{m}$ then satisfies the defining properties
of $\B{m}$ for some quantum basis $\cB=\bigcup_{m=2}^N
\B{m}$ which is extremal w.r.t. $Q_0$. But this does
{\em not} follow from the definition and is just the
point in question. In fact it is not hard to see that
\be
\dim L(\Bo{m}) =\dim\LB{m}\;,\sspace 2\leq m\leq N\;.
\ee
implies eqn~(15). In particular, once (21) has been
established, one recovers the existence of the
quantum conserved densities for $m{=}2$
\bas\nspace
&& \dim\LB{2} = \dim\LBo{2} =
               \left\{ \begin{array}{ll}
                       \#(N-1),\sspace &N-1\in E \\
                       0,        &\mbox{otherwise,}
                       \end{array}
               \right.
\eas
where $\#(N-1)$ is the multiplicity of the exponent
$N-1\in E$.
\bigskip

{\em 4.3.} In fact eqn.~(21) is stronger than eqn.~(15).
We shall see below that eqn.~(21) implies (15) for
any $g$; but (21) follows from (15) only when $g$ is
simply-laced. In particular, we claim (21) only for
simply-laced Lie algebras.

Let us first check the implication $(21)\Rightarrow
(15)$: Let $\Bo{m}=\{P^{(0)}_1,\ldots,P^{(0)}_k\}$
and $\B{m}=\{P_1,\ldots,P_k\}$ be an enumeration of
basis vectors. W.l.o.g one may assume that
$P_i^{(0)}=\Ch\, P_i$. As in the derivation of eqn.~(14)
one checks that $\Ch_{\theta}\,Q_0\, P_i =
Q_0^{(0)}\,\Ch\, P_i,\;1\leq i\leq k$. This can be done for
any $2\leq m \leq N$ and thus for a basis of $\cH^{(0)}_N$.
One concludes $\Ch_{\theta}\,Q_0=Q_0^{(0)}\,\Ch$
and hence (15).

Conversely assume now (15) to be given. For $P^{(0)}\in
\LBo{m}$ one has
$$
Q_0[\Ch^{-1}\,P^{(0)}] =\Ch^{-1}_{\theta}
[Q_0^{(0)}\,P^{(0)}] =Q_0^{(0)}\,P^{(0)} +o(\beta^2)\;.
$$
In particular for $m=2$ the argument $Q_0^{(0)}\,P^{(0)}$
vanishes (mod $L_{-1}$-exact pieces) and since
$\Ch^{-1}_{\theta}$ is injective also the $o(\beta^2)$
terms have to vanish --
which is just a rephrasing  of the conclusion following
eqn.~(15). For $m>2$ however $Q_0^{(0)}\,P^{(0)}$ is of
leading power $m-2$ and in order to confirm $\Ch^{-1}\,
P^{(0)}\in\LB{m}$ one has to show that also the quantum
corrections are of power $p\leq  m-2$. This does not follow
from (15) and requires some additional input. It turns out
that the result $iii.$ of section 2.2
provides this extra input but limits the
conclusion to simply-laced Lie algebras $g$. We shall apply
$2.2.iii.$ to $\hH_{\theta}$ and may rephrase its content
as follows: For any $R\in\hH_{\theta}$ the
set of Fock monomials $\{X_i \in\pi^{(0)}_{\theta}\}$
appearing in the expansion (8) is the {\em same} as
the one appearing in the expansion of its classical
counterpart $R^{(0)}\in\hH_{\theta}.\;(*)$ Given $(*)$
one may easily check the two inclusions
$\Ch^{-1}\LBo{m}\subset \LB{m}$ and
$\LB{m}\subset \Ch^{-1}\LBo{m}$ separately: Applied to
$R=Q_0[\Ch^{-1}P^{(0)}]\in \hH_{\theta}$,
$P^{(0)}\in \LBo{m}$ one concludes
from $(*)$ that both its classical and its quantum terms
are of power $p\leq m-2$.
Further, by construction of the classical part, no
linear combination of elements in $\Ch^{-1}\Bo{m}$
has a rest of leading power less than $m-2$ (modulo
$L_{-1}$-exact pieces) when acted upon with $Q_0$.
Hence $\Ch^{-1}\LBo{m}\subset \LB{m}$.
Conversely let $P\in\LB{m}$ be given. From $(*)$ one
concludes first that its classical part $P^{(0)}=\Ch P$
will have property (20) in the definition of $\Bo{m}$
and further that the set $\Ch\B{m}$ will satisfy condition
$b)$ in the definition whenever $\B{m}$ does. Hence
$\LB{m}\subset \Ch^{-1}\LBo{m}$. This verifies the
implication $(15) \Rightarrow (21)$ for $g$ simply-laced.
\bigskip

{\em 4.4.} In this section we introduce a system of
linear spaces displaying the relation between a basis of
type ${\cal A}$ and one of type ${\cal B}$ on $\hH_N$. The
result, eqn.~(29), can be used to compute the quantum
conserved densities. The relation between
the basis of type ${\cal A}$ and ${\cal B}$
is encoded in eqn.~(17). On the one hand it implies that
every element of $\A{m}$ has property (20) in the
definition of $\B{m}$. There may however exist linear
combinations of elements in $\A{m}$ which lie in one of
the lower spaces $\LB{p},\;p\leq m-1$, i.e.
\be
\LA{m} \subset \bigoplus_{2\leq p\leq m} \LB{p}\;.
\ee
On the other hand $\B{m}$ can contain elements of leading
power $m\leq p\leq N$ only. This is because elements of
leading power $p<m$ would by eqn.~(17) rests $R$ (modulo
$L_{-1}$-exact pieces) which are of power $m-3$ or
less -- in conflict with condition $a)$ in the definition
of $\B{m}$. Thus for $2\leq m\leq N$
\renewcommand{\theequation}{\arabic{equation}.m}
\be
\LB{m} \subset \bigoplus_{m\leq p\leq N} \LA{p}\;.
\ee
\renewcommand{\theequation}{\arabic{equation}}

For a given basis $\cB$ this condition can be
exploited recursively. For $m=N$ it reads
$\LB{N}\subset\LA{N}$, which is to say that there exists
a (non-unique) subspace $W_N\subset \LA{N}$ s.t.
\be
W_N\oplus \LB{N} =\LA{N}\;.
\ee
One can use the ambiguity in
the choice of $W_N$ to achieve that $Q_0W_N$ has a rest
of power $p\leq N-3$ (modulo $L_{-1}$-exact pieces).
By definition of $\B{N}$ there
exists a basis $\Ab{N}$ of $\LA{N}$
s.t. $\Ab{N}=\B{N}\cup V_N$, where $Q_0V_N$ has a rest
of power $p\leq N-3$ modulo $L_{-1}$-exact pieces. Thus
\be
\LV{N}\oplus \LB{N} =\LAb{N}\;.
\ee
In the next step insert (25) into the condition (23.$N-1$)
\bas
\LB{N-1}&\subset & \LA{N-1} \oplus \LAb{N} \nonum
        &   =    & \LA{N-1} \oplus \LB{N} \oplus \LV{N}\;.
\eas
By construction of $V_N$ and eqn.~(17)
this implies that in fact
\bas \nspace
&& \LB{N-1}\subset \LA{N-1} \oplus \LV{N}\;.
\eas
{}From here one concludes that there exists a (non-unique)
subspace $W_{N-1}\subset \LA{N-1}\oplus \LV{N}$ s.t.
\bas \nspace
&& W_{N-1}\oplus \LB{N-1}= \LA{N-1} \oplus \LV{N}\;.
\eas
Again we exploit the ambiguity by chosing a basis
$\Ab{N-1}\cup \Vb{N}$ of $\LA{N-1}\oplus \LV{N}$ s.t.
$\Ab{N-1}\cup \Vb{N} =\B{N-1} \cup V_{N-1}$, where
$Q_0V_{N-1}$ has a rest of power $p\leq N-4$ modulo
$L_{-1}$-exact pieces. Thus
\be
\LV{N-1}\oplus \LB{N-1}= \LAb{N-1} \oplus \LVb{N}\;.
\ee
By induction on $m$ one can now infer the existence of
subspaces $W_m\subset \LA{m} \oplus \LV{m+1}$ s.t.
for $3\leq m\leq N$
\renewcommand{\theequation}{\arabic{equation}.m}
\be
W_m\oplus \LB{m}= \LA{m} \oplus \LV{m+1}\;.
\ee
\renewcommand{\theequation}{\arabic{equation}}
and $Q_0V_{m+1}$ has a rest of power $p\leq m-2$ modulo
$L_{-1}$-exact pieces.
We include (24) in the chain of equations (27.m) by
defining $V_{N+1}=\emptyset$. For $m<N$ the equation
(27.m) is obtained from the condition (23.m) applied to
the basis
\bas \nspace
&& \cA(m+1) := \left(\bigcup_{p=m+1}^N\Ab{m}\right) \cup
               \left(\bigcup_{p=2}^m \A{m}\right)\;.
\eas
By inserting consecutively the equations (28.p) below,
$p=N,\ldots ,m+1$ and simplifying the resulting
inclusions one arrives at
$\LB{m}\subset\LA{m} \oplus \LV{m+1})$, from
which one concludes (27.m). The induction step is driven
by the choice of a basis $\Ab{m} \cup \Vb{m+1}$ in
$\LA{m}\oplus \LV{m+1}$ s.t. $\Ab{m} \cup \Vb{m+1}=
\B{m} \cup V_{m}$, where $Q_0V_{m}$ and
$Q_0\Vb{m+1}$ have a rest of power $p\leq m-3$ and
$p\leq m-2$, respectively, modulo
$L_{-1}$-exact pieces. The equation (27.m) becomes
\renewcommand{\theequation}{\arabic{equation}.m}
\be
\LV{m}\oplus \LB{m}= \LAb{m} \oplus \LVb{m+1}\;,
\ee
\renewcommand{\theequation}{\arabic{equation}}
for $3\leq m\leq N$. In particular, the basis
$\cA(m+1)$ has been transformed into $\cA(m)$,
completing the induction step. The last equation (28.3)
reads $\LV{3}\oplus \LB{3}=\LAb{3}\oplus \LVb{4}$,
where by construction $Q_0V_3$ and $Q_0\Vb{4}$ have
rests of power $p=0$ and $p\leq 1$, respectively.
But this means that $\LVb{3}$ actually is the space
of conserved densities $\LB{2}$ and we may assume
that $V_3=\B{2}$. One can also check that the subsequent
equation (22.8) would just reproduce this feature:
For $N>2$ both $V_2$ and $\Ab{2}$ are empty, so that
(22.8) collapses to $\LB{2}=\LVb{3}$.
In summary we have shown:
\bigskip

\noindent
For a given basis $\cB,\;N>2$ there exists a basis $\cAb :=
\cA(2)=\bigcup_{m=2}^N\Ab{m}$ and vector spaces
$\LV{m},\;\LVb{m},\;3\leq m \leq N$ s.t.
\begin{itemize}
\item[--] $\dim\LV{m} =\dim\LVb{m}\;.$
\item[--] $\LV{m}$ and $\LVb{m}$ contain elements of
leading power $p\geq m$ only.
\item[--] $Q_0V_m$ and $Q_0\Vb{m}$ have rests of power
$p\leq m-3$ modulo $L_{-1}$-exact pieces.
\end{itemize}
In particular $\LV{3} =\LVb{3} =\LB{2}$.
The relations among these spaces are summarized in the
tableau
\ba
\;\;\;& \;\;& \LV{N}\oplus \LB{N} =\LAb{N} \nonum
\;\;\;&     & \LV{m}\oplus \LB{m}=\LAb{m}\oplus \LVb{m+1}\;,
              \bspace 4\leq m\leq N-1 \nonum
\;\;\;&     & \LB{2}\oplus \LB{3}= \LAb{3}\oplus \LVb{4}\;.
\ea
These features follow directly from the construction
outlined. In particular, the equation
$\dim\LV{m} =\dim \LVb{m}$ follows from $\dim\LA{m}
=\dim\LAb{m}$ and guarantees that the dimensions
on both sides of (29) add up correctly:
$\sum_{m=1}^N\dim\,\LB{m}= \sum_{m=1}^N\dim\,\LA{m}$,
where both sides equal $\dim\,\hH_N$. We repeat that all of
the spaces $\LV{m}\;,\LVb{m},\;\LAb{m-1},\;4\leq m\leq N$
depend on the initial choice of basis $\cB$ and only their
dimensions are uniquely defined. The exceptional cases are
$\LV{3} =\LVb{3}=\LB{2}$ where already the vector spaces
(not only their dimensions) do not depend on the choice
of basis. In addition we have
\be
\dim\LV{N} \geq \dim\LV{3} =\dim\LB{2}\;.
\ee
To see this, recall that quantum conserved densities
have to be of leading power $N$ in $\{x_n^i\}$. Clearly
$\LB{N}$ is useless for the construction of conserved
densities, so that $\dim\LV{N} \geq \dim\LV{3}$.
\bigskip

{\em 4.5.} Now consider the relation between the tableau (29)
and its classical counterpart, where the sets
$\A{m},\;\B{m},\;V_m,\;\Vb{m}$ are replaced by
$\Ao{m},\;\Bo{m},\;V_m^{(0)},\;\Vbo{m}$, respectively.
We will denote this tableau by $(29)^{(0)}$. The sets
$\Bo{m},\;2\leq m\leq N$ have already been defined.
Let ${\cal A}_N^{(0)}=\bigcup_{2\leq m\leq N} \Ao{m}$
denote the classical counterpart of $\cA$ i.e. a basis
of $\hH_N^{(0)}$ having properties $a)$ and $b)$ in
the definition. Since eqn.~(17) is preserved in the classical
case, the basis ${\cal A}_N^{(0)}$ and ${\cal B}_N^{(0)}$
are related  by a chain of equations $(23.m)^{(0)},\;
2\leq m\leq N$, replacing $(23.m)$. Given these, the derivation
of the classical counterpart of the tableau (29) can be done
as before. In particular, this defines the subspaces
$L(V_m^{(0)})$ and $\LVbo{m}$. The form of the classical tableau
$(29)^{(0)}$ will be the same as (29), but in principle
the dimensions of some of the linear spaces might be
different. However, this is not the case.

Since the subspaces $\LAo{m}$ are defined by means
of properties of the leading power terms only (which are
left invariant by $\Ch$), they manifestly have the same
dimensions as their classical counterparts
\be
\dim\LAo{m} =\dim\LA{m}\;,\sspace
2\leq m\leq N\;,
\ee
and one may assume w.l.o.g. that $\Ao{m}=\Ch \A{m}$.
For other subspaces $U\subset \hH_N$, which are defined
through conditions that affect also subleading powers,
a relation of the form (31) is not manifest -- which is
just why the equation (21) required proof. Given that both
$\LA{m}$ and $\LB{m}$ have the same dimensions as their
classical counterparts, the same follows for the remaining
spaces $\LV{m},\;\LVb{m},\;3\leq m \leq N$. This is easily
verified by induction from top to bottom and right to left
in the tableau (29).
\bigskip

{\em 4.6.} The tableau (29) gives a means to construct the
quantum conserved densities. The method is to calculate the
sets $V_N,\Vb{N},\;V_{N-1},\Vb{N-1},\;\ldots,
V_3,\Vb{3}$ recursively. In the final step one obtains a
basis $V_3=\Vb{3}=\B{2}$ for the space of quantum
conserved densities at degree $N$. Moreover, from the
previous results, all of the states invoked are in 1-1
correspondence to their classical counterparts. The
calculation in the classical and the quantum case
therefore is principally the same -- just the coefficients
in the linear combinations of the basis vectors in $\cA$
get deformed.

In detail, suppose a basis of type ${\cal A}$ to be given.
For example one may take the set of states
(11) of degree $N$. Consider now the top row
of (29). From (30) one sees that a necessary condition
for a conserved density to exist at degree $N$ is that
$V_N$ is nonempty. For $N-1\in E$ the elements of
$V_N$ therefore are the initial material for
the construction of the conserved densities.
In the second row of (29) one adds the basis elements
of $\LA{N-1}$ and eliminates the space of linear
combinations $\LB{N-1}$ which is certainly useless
for the construction of the conserved densities. The
remainder $V_{N-2}$ is used as an input for the next
row etc.. The result $\dim\LV{N}\geq \dim\LB{2}=\#(N-1)$
means that for $N-1\in E$ this process iterates and
in the last step yields a basis $V_3=\B{2}$ for the
space of conserved densities.

Examples for such  calculations for the low rank $sl_n$-cases
can be found in \cite{MN1}. The above scheme
(from a slightly different viewpoint) was referred to
as `elimination algorithm'. In these cases it appeared that
in fact all the spaces $\LV{m}$ are non-trivial and have the
same dimension $\dim \LV{m} =\#(N-1),\;2\leq m\leq N$. We
expect this to be correct in general.
%%%%%%%%%%%%%%%%%%%%%%%%%%%%%%%%%%%%%%%%%%%%%%%%%%%%%%%%%%%
\newsection{Conclusions}
The classical vertex operator $Q_0^{(0)}$ and the quantum
vertex operator $Q_0$ were shown to be related by an
equivalence transformation. This implies that a
basis $\cB$ containing the conserved densities as some of
the basis vectors can smoothly be deformed from the classical
to the quantum theory. This class of basis can be characterized
by certain extremal properties which give rise to a
construction scheme for both the classical and the quantum
conserved densities. The results yield a novel understanding
of the role of the quantization process in $W$-algebras.
A better control of the transformation (15) would give a
means to quantise many concepts of the classical integrable
hierarchies in a systematic fashion.

\vspace{1cm} \noindent
{\tt Acknowledgements:} I wish to thank E. Frenkel
for stimulating discussions and for contributing
to the clarity of the argument.
%
%%%%%%%%%%%%%%%%%%%%%%%%%%%%%%%%%%%%%%%%%%%%%%%%%%%%%%%%%%%

%%%%%%%%%%%%%%%%%%%%%%%%%%%%%%%%%%%%%%%%%%%%%%%%%%%%%%%%%%%%
%
\end{document}